Opinion Instability and Measurement Errors:
A G-Theory Analysis of College Students

Bang Quan Zheng

School of Government and Public Policy
University of Arizona

2024

Abstract

This study examines opinion instability among individuals from different ethnic groups (White, Latino, and Asian Americans) by analyzing measurement errors in survey measures. Using a multi-wave panel dataset of college students and employing generalizability theory, the study uncovers significant patterns. The results reveal that White students exhibit higher attitude reliability, characterized by larger variances in true opinions and smaller measurement errors. In contrast, Latino and Asian American students display lower attitude stability, with lower variances in true opinions and higher variances in both item-specific and measurement errors. Disparities in political socialization and issue concerns contribute to the observed attitude instability among Latino and Asian American students. Moreover, Asian American and Latino respondents require a greater number of survey items to mitigate measurement error compared to their White counterparts. However, the impact of multiple waves of surveys on improving reliability is limited for Latino and Asian American students compared to White students. These findings deepen our understanding of attitude stability across ethnic groups and underscore the importance of further research in this area.

[Word Count: 8,486]

# 1 Introduction

Political representation in democratic theory assumes that ordinary individuals possess well-informed attitudes on major public policy issues, and public opinion serves as the primary mechanism for gauging the preferences of the general public (Ansolabehere et al., 2008; Downs, 1957; Key, 1966). However, existing studies in public opinion research suggest that ordinary individuals often hold changeable preferences on political issues and their responses to survey questions can contain a significant element of randomness (Jackson, 1983; Judd and Milburn, 1980; Moskowitz and Jenkins, 2004; Norpoth and Lodge, 1985). Measurement error can affect the validity of survey results and their statistical significance (Achen, 1983; Feldman, 1989). Therefore, both the malleability of public opinion and the presence of measurement error are significant factors to consider. In order to ensure accurate measurement of public opinion, it is essential to comprehensively understand the sources of measurement error and implement strategies to reduce its impact. Despite being a long-acknowledged issue, survey measurement problems are seldom examined in public opinion research. As scholars have pointed out, survey items alone may not sufficiently measure the variation in broad predispositions that drive individuals' preferences (Pietryka and MacIntosh, 2022).

This survey measurement issue carries particular significance when examining the mass opinions of racial minorities. Existing studies on public opinion have predominantly focused on native-born White Americans, often assuming that minority individuals experience similar measurement error in surveys or neglecting their distinctive characteristics altogether, despite the changing demographic landscape of the United States. With Latino and Asian American populations increasingly influencing electoral outcomes (Hajnal and Lee, 2011; Wong et al., 2011), any attempt to understand the public opinion of these groups must account for their unique characteristics in survey responses and measurement metrics.

This paper aims to investigate the sources of measurement error and uncover divergent patterns of group-specific variances in opinion reliability among White, Asian American, and Latino participants. The study focuses on decomposing the measurement error into two components: item-specific variance and over-time specific variance. By analyzing these components, a deeper understanding of the factors influencing opinion reliability within different groups can be achieved. Despite being a longstanding concern in public opinion research, this issue has received limited empirical attention. Specifically, our study examines two critical sources of measurement error: temporal instability and item-based instability, which can be attributed to the expression of attitudes (Zaller and Feldman, 1992).

The metric for measuring opinion stability involves calculating the ratio of true opinion variance, or the variance of the signal, to the total variance. With this measure, there are two primary ways of conceptualizing opinion instability in survey measures. One view posits that instability stems from inconsistencies in individuals' true opinions. This perspective holds that individuals hold partial considerations on policy issues, which results in fickle survey responses when sampling from these pools of inconsistent considerations. The cause of these inconsistencies can be attributed to varying levels of political knowledge (Zaller, 1992; Zaller and Feldman, 1992). In contrast, another view argues that opinion instability arises from measurement error. In this view, individuals have coherent issue preferences and underlying attitudes that are driven by predispositions such as core values. Here, low consistency in attitudes is attributed to measurement error (Achen, 1975; Ansolabehere et al., 2008; Loken and Gelman, 2017). These studies share the belief that item-specificity, or preformed attitudes at the level of specificity demanded in survey items, has important implications for measurement validity and reliability. Therefore, our understanding of opinion reliability largely hinges on our ability to comprehend the random variance components in survey responses (Feldman, 1989).

A growing body of scholarship in public opinion has revealed that ordinary individuals hold poorly formed attitudes on many political issues, while measurement error is typically treated as random noise. To address these concerns, we utilize generalizability theory, also known as G theory, to assess item-based and temporal instability across distinct racial groups. Our findings suggest that the degree of response instability varies not only across different policy domains but also across racial categories. Specifically, we identify two salient patterns. First, White participants display a larger true opinion variance and a smaller error variance, thus enabling higher attitude reliability. By contrast, Latino and Asian American respondents exhibit lower attitude stability, constrained by low true opinion variance and high item-specific variance. Hence, our paper contends that the mix of partially consistent ideas and considerations in item-specificity impacts the attitude stability of Latino and Asian American students. Moreover, our study highlights that Asian American and Latino participants require a greater number of survey items to reduce measurement error compared to their White counterparts.

Political socialization is a crucial component in the development of coherent political knowledge, attitudes, values, and participation among university students (Campbell et al., 1960; Ehman, 1980). However, the racial differences in political socialization can lead to different opinions and survey response characteristics. Although using student samples for research has limitations, they possess desirable characteristics such as unique patterns of response randomness. The controlled and educative social environment of university campuses, coupled with the students' higher education levels, render them ideal comparison groups in longitudinal studies. Previous studies have shown that political socialization within schools presents an excellent opportunity for observing the formation of political attitudes and stability (Laar et al., 2005).

This article argues that the presence of extraneous factors in item-specificity renders Latino and Asian American respondents more prone to a combination of partially consistent considerations in answering survey questions compared to White respondents. The low opinion stability of Latinos and Asian Americans is attributed to the interplay between low true opinion variance and high item-specific variance. Conversely, native-born White Americans, as a collective, exhibit larger variance in true opinions and smaller item-specific variance. These results are drawn from the five-wave panel data from UCLA intergroup conflict data spanning from 1996 to 2000. Despite the possibility that the data may be dated, this dataset remains the only multiple wave panel dataset that provides a comprehensive range of variables suitable for conducting cross-group analysis and examining political socialization in college. By utilizing G theory, the study disentangles error variance and contrasts these error variance components across White, Asian, and Latino students. As demonstrated by this research, the G theory model offers a more comprehensive explanation of opinion instability and provides a more precise measurement of opinion reliability. The study reveals that White students tend to align with two polarized partisan coalitions, thus possessing considerable opinion variance and higher over-time variability but smaller item-specific variance. These traits make it more straightforward to decrease the overall error variance and enhance the measurement's reliability. In contrast, Latino students, on the whole, strongly identify with the Democratic Party, resulting in smaller true opinion variance that correlates with the items but larger item-specific variance, leading to challenges in attenuating error variance. Asian Americans fall in between Whites and Latinos, with low true opinion variance and moderate item-specific variance.

    This paper addresses the divergence of arguments on opinion instability by incorporating group-based characteristics, examining measurement error sources in survey research, particularly regarding group differences. It emphasizes the impact of measurement error in survey responses from minority individuals, compromising measurement validity and capturing unintended aspects. The findings derived from G theory shed light on these measurement error sources and offer valuable insights for public opinion research, enabling the reduction of measurement error and enhancing opinion reliability. Due to the limited availability of longitudinal data spanning multiple waves, there is a

scarcity of empirical studies investigating sources of political attitude instability. Thus, this study on opinion stability represents a significant step towards bridging this gap in the literature.

The paper is organized as follows: In the second section, we provide a literature review on instability in public opinion. Then, in the third section, we discuss the multi-item scale and G theory. The fourth section introduces the data and methodology used in this study. Descriptive statistics, including key variables and political socialization across different groups, are presented in the fifth section. Moving on to the sixth section, we delve into the analysis of policy attitude items, examining their determinants and dimensionalities across groups. In the seventh section, we present the statistical results derived from G theory. Finally, the last section encompasses a comprehensive discussion and conclusion.

## 2   Previous Studies on Opinion Instability

There exist two primary theoretical strands regarding the instability of survey measures of individuals' opinions. The first strand contends that instability in survey measures arises from true opinions. Political conceptualization is a significant factor in this perspective, as it reflects the ideological or partisan bundles among parties, policy views, and social identities (Converse, 1964). However, such conceptualization varies among the general public, as only a small proportion of individuals can interpret political behavior or issues with coherent ideological considerations (Campbell et al., 1960) or make sophisticated use of political abstraction (D. Kinder, 1998; Zaller, 1992).

Politically sophisticated individuals are expected to have knowledge of the connections between parties, candidates, and policy issues, which allows them to use party or candidate positions as heuristics (Bullock, 2011; Norpoth and Lodge, 1985; Zaller, 1992). In contrast, most ordinary people only possess partial or inconsistent considerations on political issues, with their political thinking relying on group benefits, the information to which they are exposed, or their mood when taking a survey. Sampling from these pools of inconsistent considerations, their survey responses tend to be fickle (Norpoth and Lodge, 1985; Zaller, 1992; Zaller and Feldman, 1992).

Additionally, scholars in this field of theory argue that having knowledge about the topic presented in a survey is necessary for stability, and differences in stability are a result of discrepancies in underlying political knowledge or concerns. This view, known as item-specificity, suggests that opinion variability can be attributed to variations in political knowledge or issue concerns. In support of this, Freeder et al. (2018) posit that understanding "what goes with what" is crucial for attitude stability. Individuals often align their views with the policy positions of their preferred parties (Bullock, 2011; Lenz, 2012). As people acquire knowledge of party policy positions, they are more likely to exhibit stable policy views (Freeder et al., 2018). Based on this logic, item-specificity should exhibit high temporal variability and low variability across items within the same issue areas because people are expected to learn political bundles over time. To test this hypothesis, it is necessary to examine the time-varying changes in repeated measurements.

On the contrary, some scholars suggest that there exists a consistent underlying preference and an additive random error in survey responses (Ansolabehere et al., 2008). They argue that a set of information, values, and beliefs forms an individual's true opinions (Alvarez and Brehm, 2002; Feldman, 1989; Zaller, 1992). As a result, true attitudes are expected to be coherent across repeated measurements, and only the random error changes. When policy issues are framed in a manner relevant to an individual's core values, respondents can easily make a choice (Alvarez and Brehm, 2002; Feldman, 1989; McClosky and Zaller, 1984; Zaller, 1992). Furthermore, the salience and involvement of an issue indirectly lead to less response instability (Feldman, 1989; D. Kinder, 1983). Therefore, differences in measurement precision across attitude object categories are attributed to the respondents' interpretation of survey questions rather than the attitude object categories themselves (Krosnick, 1991).

Item-specificity is closely related to the levels of relevant political conception, which may vary across different demographic groups. Asian Americans and Latinos tend to conceptualize political issues differently from their Anglo counterparts due to factors such as incomplete information, ambivalent ideology, and uncertain identity (Abrajano and Alvarez, 2011; Hajnal and Lee, 2011). As a result, certain survey questions may be understood differently in various groups and may vary across different contexts (Perez and Hetherington, 2014).

Compared to native-born White Americans, a significant percentage of Asian Americans and Latinos were born outside the United States, which leads to weaker political socialization (Carlos, 2018; Hajnal and Lee, 2011; Raychaudhuri, 2018). While White Americans tend to learn about politics and develop partisanship from their parents, communities, and peers (Campbell et al., 1960; Jennings et al., 2009; Sears and Funk, 1999), many Asian Americans learn about politics mostly from peers and mass media rather than their parents (Wong, 2000; Wong and Tseng, 2007). Consequently, many Latinos and Asian Americans possess uncertain and ambivalent attitudes towards political issues, leading them to answer "don't know" in surveys (Hajnal and Lee, 2011). Nonetheless, it remains a mystery as to how these differences in political socialization translate into opinion reliability.

However, the distinction between true opinion variance and error variance hinges on the level of item-specificity. While item-specificity is crucial, its root cause remains unclear and is beyond the scope of this study. It is unclear whether time-varying repeated measurements or content-specificity causes item-specific variance. Existing studies on opinion instability only focus on one source of measurement error at a time, despite time-varying occasion-specific variance in repeated measurements and item-specific variance being the two most common sources of measurement error. If individuals are asked the same survey questions in a series of repeated interviews, their time-varying opinion reports across different measurements could be inconsistent due to various random factors, such as vague question wordings, ambiguous response categories that do not include the respondent's preference (Eady, 2017; Mosteller, 1968), or the language used during the interview (Lee and Perez, 2014). Moreover, existing studies tend to overlook the underlying predispositions of minority citizens and their impact on public opinion measures. Researchers assume that survey items or scale measures within one group are equivalent to those of another group (Pietryka and MacIntosh, 2013).

## 3 Attitude Stability Measurement and G Theory

In this section, we briefly review the measurement of attitude stability, emphasizing the limitations of the multi-item scale method and introducing the concept of G theory. While the multi-item scale is a commonly employed and standardized method for assessing reliability, G theory offers unique advantages. It can identify and account for multiple sources of measurement error concurrently, making it particularly suitable for capturing individual, temporal, and item-specific variances that extend beyond the capabilities of the multi-item scale. Specific distinctions between these methods are illustrated in the subsequent pages.

### 3.1 Multi-item Scale

The standard measure of stability of public opinions is based on the ratio of true attitude variance to error variance. Theoretical models of measurement error in surveys treat responses to individual questions, or items, as comprising the true attitude and random error. Specifically, if $X$ represents the observed response on a given item, it can be decomposed into $X = T + E$, where $T$ denotes true attitude and $E$ denotes random error. The population variance can then be decomposed as follows:

$$\sigma_X^2 = \sigma_T^2 + \sigma_E^2 + \sigma_{T,E}^2 , \tag{1}$$

Where $\sigma_T^2$ represents true opinion variance, $\sigma_E^2$ represents error variance, and $\sigma_{T,E}^2$ denotes the interaction between true opinion and error variance. In general, true opinion variance $\sigma_T^2$ is assumed to be constant and uncorrelated with error variance $\sigma_E^2$, resulting in $\sigma_{T,E}^2 = 0$. Thus, we can simplify equation 1 to:

$$\sigma_X^2 = \sigma_T^2 + \sigma_E^2. \qquad (2)$$

The reliability of opinions can be measured by the ratio of the true opinion variance to the total variances, denoted by $\rho_{XX}$. For multiple items, the reliability is calculated as:

$$\rho_{XX} = \frac{\sigma_T^2}{\sigma_X^2}, \qquad (3)$$

where $\sigma_X^2$ is the variance of the observed scores. By substituting $\sigma_X^2$ with $\sigma_T^2 + \sigma_E^2$ (as shown in equation 2), we can obtain:

$$\rho_{XX} = \frac{\sigma_T^2}{\sigma_T^2 + \sigma_E^2}. \qquad (4)$$

which measures the reliability of multiple items.

Previous research has shown that averaging multiple items can help minimize the error variance $\sigma_E^2$ and increase reliability (Ansolabehere et al., 2008; Freeder et al., 2018). Equation 5 demonstrates this logic, where $n_i$ represents the number of items:

$$\rho_{XX} = \frac{\sigma_T^2}{\sigma_T^2 + \frac{\sigma_E^2}{n_i}}. \qquad (5)$$

As the number of items increases, the error variance $\sigma_E^2$ decreases, resulting in higher reliability. However, equation 5 treats all measurement errors as one source and cannot separate error variance into different sources to measure their specific effects. According to Zaller (2012), correcting for the aggregate measurement error fails to distinguish between sources of random variability and only corrects for all sources of measurement error.

## *3.2 G Theory and Measurement Error Attenuation*

From a measurement perspective, various potential sources of measurement error can impact the estimation of opinion stability. G theory is a statistical approach that decomposes measurement error to calculate reliability, allowing for a comprehensive analysis of the measurement errors generated by different sources. These errors may stem from inconsistencies in item interpretation, changes in partisanship over time, or variations at different points during repeated surveys. While a standard multi-item scale can be utilized to estimate opinion stability, it is unable to account for multiple sources of error simultaneously. The multi-item scale measures the aggregate level reliability statistic, which Jones and Norrander (1996) note is assessed in terms of aggregate units, rather than individual responses.

G Theory decomposes variance components, enabling us to pinpoint sources of systematic and unsystematic error variation and estimate each possible combination of the interactions between them (Brennan, 2001; Shavelson and Webb, 1991). For instance, in multi-wave panel survey data, the sources of randomness may stem from individual variation $\sigma_p^2$, survey item interpretation $\sigma_i^2$, measurement occasions between repeated surveys $\sigma_o^2$, the interaction between individual and survey item $\sigma_{pi}^2$, as well as the interaction between individual, item, occasion and all other variances $\sigma_{pio,e}^2$. Thus, the variance components are $\sigma_p^2$, $\sigma_i^2$, $\sigma_o^2$, $\sigma_{pi}^2$, and $\sigma_{pio,e}^2$, and the total variance is the weighted sum of these components.

$$E\rho^2 = \frac{\hat{\sigma}_p^2}{\hat{\sigma}_p^2 + \frac{\hat{\sigma}_{pi}^2}{n_i'} + \frac{\hat{\sigma}_{po}^2}{n_o'} + \frac{\hat{\sigma}_{poi,e}^2}{n_i' n_o'}}, \qquad (6)$$

The generalizability coefficient or reliability of internal consistency, denoted as $E\rho^2$, is calculated using equation 6. This coefficient ranges from 0 to 1, where higher values indicate higher reliability. The equation includes terms for the variances of persons, items, and measurement occasions, as well as their interactions, which are denoted by $\hat{\sigma}_p^2$, $\hat{\sigma}_{pi}^2$, $\hat{\sigma}_{po}^2$, and $\hat{\sigma}_{pio,e}^2$, respectively. The numbers of items and measurement occasions are denoted by $n_i'$ and $n_o'$, respectively. The interactions between person and item variance and between person and measurement occasion variance are attenuated by $\frac{\hat{\sigma}_{pi}^2}{n_i'}$ and $\frac{\hat{\sigma}_{po}^2}{n_o'}$, respectively. This allows us to measure the amount of error variance specifically generated by the interactions between true opinion and items and between true opinion and measurement occasions. The interaction between person, item, occasion, and all other variances are attenuated by $\frac{\hat{\sigma}_{pio,e}^2}{n_i' n_o'}$, which accounts for all other sources of error variance.

As shown in Equation 6, the strength of G theory lies in its ability to estimate multiple sources of error variance in a single model. It can consider interactions between the variances of persons and items, between item-specific and occasion-specific variances, and between the variances of persons and measurement occasions, as well as the 3-way interaction between the variances of items, occasions, and residuals.

## 4 Data and Method

This study utilized data from the UCLA Intergroup Student Conflict Studies (1996-2000), a longitudinal study tracking incoming first-year students at UCLA over five years. The dataset included repeated surveys administered annually during this period. The incoming first-year class consisted of 3,877 students, with 32 percent White, 36 percent Asian American, 18 percent Latino, 6 percent African American, and 8 percent belonging to another ethnicity or not reporting. Data collection occurred during five different time periods between 1996 and 2000, with the first wave collected during the summer orientation program in 1996. Subsequent data was obtained during the spring quarter of each academic year from 1997 to 2000, though there was a decline in respondents completing all waves. African Americans were excluded from the study due to a small sample size across the five waves. Statistical analysis employed the GENOVA software, focusing on policy attitude issues measured by eight items in the panel data. The interviews were conducted using the Computer-Assisted Telephone Interview (CATI) system, operated by the Institute for Social Science Research at UCLA, with an average duration of 20 minutes per interview.

The statistical analyses undertaken in this study encompass a range of approaches, including regression analysis and the more advanced G theory approach, providing incremental and diverse perspectives. Initially, we employ regression analyses to examine the relationship between policy attitude items, party identification, and identity variables. This regression analysis serves as an initial step to identify the potential differing determinants of the policy attitude items among White, Latino, and Asian American students. Next, we utilize factor analysis to explore the multi-dimensionality that underlies the policy items across these distinct groups.

Subsequently, we apply G theory to investigate the sources of instability in policy attitudes and attenuate the measurement errors. In the context of G theory, the focus of measurement is on students, with the two facets being survey items (I) and waves of measurement occasions (O). This study assumes that the student samples were randomly and independently drawn, and the effects of survey items and repeated measurements were also independently and identically distributed. Therefore, the variance components are $\sigma_p^2$, $\sigma_i^2$, $\sigma_o^2$, $\sigma_{pi}^2$, and $\sigma_{pio,e}^2$. Additionally, this study employs a fully crossed random effect design, which allows for interactions of all sources of error variances to avoid confounding effects (Brennan, 2001; Shavelson and Webb, 1991).

## 5 Demographic Characteristics and Political Socialization

In this section, we will delve into the demographic characteristics of White, Latino, and Asian American students and their relationship to general cognitive abilities, partisan orientation, and political ideology. Examining these characteristics is crucial as they help determine whether individuals possess the necessary framework through which they interpret the policy attitude items. Political socialization can lead to variations in policy attitudes and survey response patterns. This is especially significant when comparing US-born White Americans to Asian Americans and Latinos, as these two groups tend to have weaker pre-adult political socialization (Hajnal and Lee, 2011). Among the student subjects in this study, 75.1 percent were born in the United States. In terms of group levels, 94.5 percent of White students, 47.7 percent of Asian American students, and 85.5 percent of Latino students were born in the United States. Despite approximately half of the Asian sample being foreign-born, a mere six percent of them were identified as international students.

Cognitive abilities can affect opinion stability when interpreting survey questions. In particular, a large proportion of foreign-born Asian Americans and Latinos have limited English skills or poor political cognition, which could be why they tend to answer "don't know" in survey responses (Kim and Lee, 2001). Using student samples can address this concern, as we can use their SAT scores in the verbal and math parts to assess their cognitive competence. Figure 1 shows the mean scores of SAT in the verbal and math sections, along with 95 percent confidence intervals. As shown, White students tend to have higher SAT scores in the verbal section, while Asian and Latino students tend to have similar performances. For SAT math scores, Asian students outperformed other groups. However, these overall differences do not seem to have a significant impact on the interpretation of survey questions.

Figure 1. Mean SAT Scores

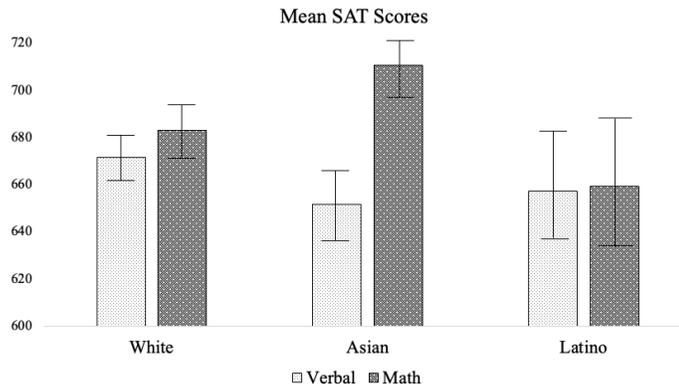

Partisanship is a crucial factor that shapes individuals' policy preferences and attitude dynamics (Campbell et al., 1960; Franklin and Jackson, 1983; Jacoby, 1988). Moreover, different groups tend to have unique patterns for the acquisition of partisanship. University campuses play a vital role in providing an environment for students to socialize their partisan preferences and political attitudes.

Figure 2. Over-time partisanship across groups

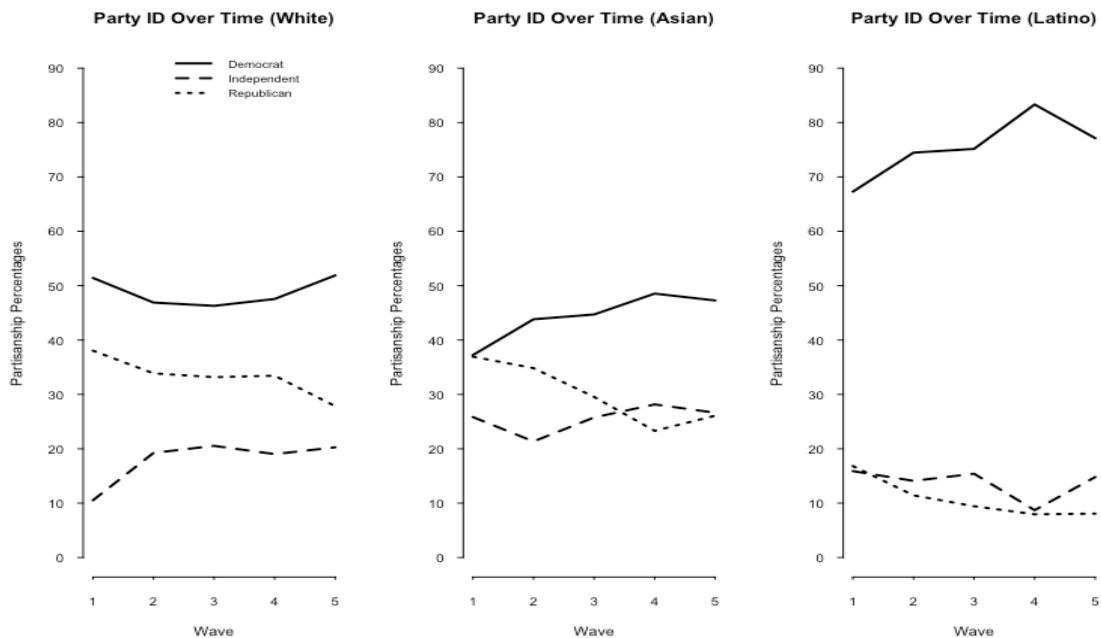

Figure 2 displays the average proportions of Democrats, Independents, and Republicans over the panel data. The first plot shows the partisan dynamics of White students. As shown, the partisanship of White students remained relatively stable, with the only noticeable change being the decrease of Republicans and the increase of Independents over time. When White students started college, approximately 38 percent identified as Republicans; five years later, this number dropped to 28 percent. Conversely, the proportion of Independents increased from 10 percent to 20 percent.

For Asian American students, the pattern was different. Approximately 38 percent of Asian Americans identified with Democrats when they started college, and this number increased to about 50 percent after five years. In contrast, the proportions of Independents remained relatively stable. Latino students had the highest proportion of Democrats when they started college, and this number increased from 69 percent to about 83 percent in four years, dipping slightly in their fifth year. In sharp contrast, only about 17 percent of Latino students identified as Republicans in their first year of college, and this number decreased to 10 percent in their fifth year. In summary, the proportion of Asian American and Latino Democrats increased by about 13 percent during their time in college, while the proportion of those who identified with Republicans decreased by 10 percent.

## 5.1 *Partisanship and Ideology Stability*

Partisanship and political ideology consistently emerge as the most robust predictors of general political and policy attitudes in American politics. The greater stabilities of partisan identification and ideology indicate a heightened level of political socialization (Campbell et al., 1960; Green et al., 2002; Lewis-Beck et al., 2008; Newcomb et al., 1965). While the college experience holds significant importance in the political socialization process, its impact varies among different groups. Research indicates that White students tend to have earlier exposure to politics compared to minority groups, attributed to their pre-adult political socialization. Many White students have already established stable political views during high school (Dawson and Prewitt, 1969; Sears and Funk, 1999). On the other hand, institutions are more influential in shaping the partisanship development of Latino and Asian American students, as individuals often adopt the opinions of the majority to fit in (Carlos, 2018; Sinclair, 2012). As a result, the college experience for Latino and Asian American students can be seen as an extended political socialization process, during which they are exposed to different ideas and standard political debates of American politics. Therefore, it is expected that White students would exhibit a stronger internal consistency of self-reported partisanship and political ideology compared to their Asian American and Latino counterparts.

Figure 3. Partisanship and Ideology reliability across waves

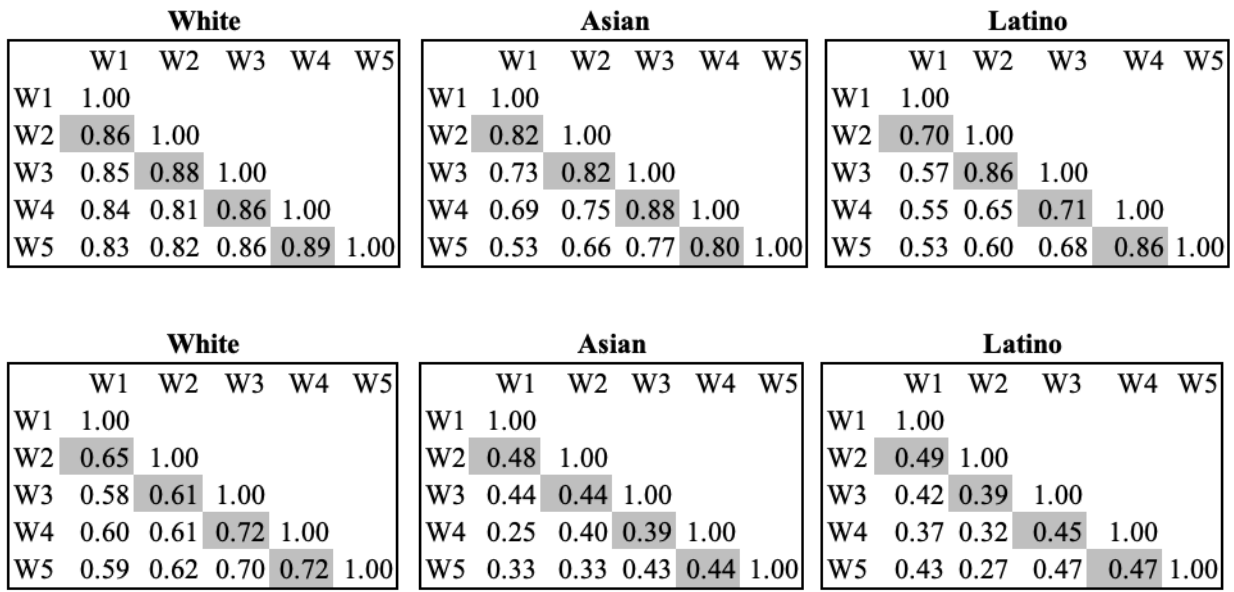

To what extent did individuals maintain consistency in their partisanship and ideology over time? The top panels of Figure 3 display correlation matrices depicting the relationships between partisanship across five waves of surveys. The findings demonstrate strong and consistent correlations among White students, with the correlation increasing from 0.86 to 0.89 from the first to the last wave. Notably, these correlations consistently remained above 0.8, indicating a high degree of stability in partisanship among White students. Similarly, Asian American students exhibited relatively stable correlations, ranging from 0.80 to 0.88 across the five waves. In contrast, Latino students displayed more variability, with correlations ranging from 0.70 to 0.86 over the same period. The presence of lower reliabilities in party identification and political ideology suggests the existence of measurement errors among Latino and Asian American respondents. This highlights the potential influence of measurement inconsistencies on the observed variability in correlations for these groups.

## 6   Policy Attitude Items

In this section, we introduce the eight carefully selected policy attitude items that are crucial for examining item-specific variances and measurement errors. These items encompass broader social attitudes that hold significant relevance for racial minority and immigrant groups, making them particularly pertinent to Latinos and Asian Americans. Additionally, apart from education and political knowledge, the stability of opinions can also be influenced by the salience and level of involvement individuals have in the issues at hand (Converse, 1964). According to scholars, minorities tend to support policy issues that benefit their own groups or those with whom they share social status (Campbell et al., 1960; D. R. Kinder and Kam, 2009; Levin and Sidanius, 1999; Sidanius and Pratto, 1999). Moreover, when it comes to racial predispositions, people are more likely to stick to their attitudes than to adjust their opinions to align with their preferred political parties, unlike in economic policy issues (Tesler, 2015). Therefore, these attitude items enable us to explore attitude dynamics across different groups and over time, from 1996 to 2000. (Refer to Figure A2 for the item distributions of the various groups).

Figure 4. Policy attituden trends over time

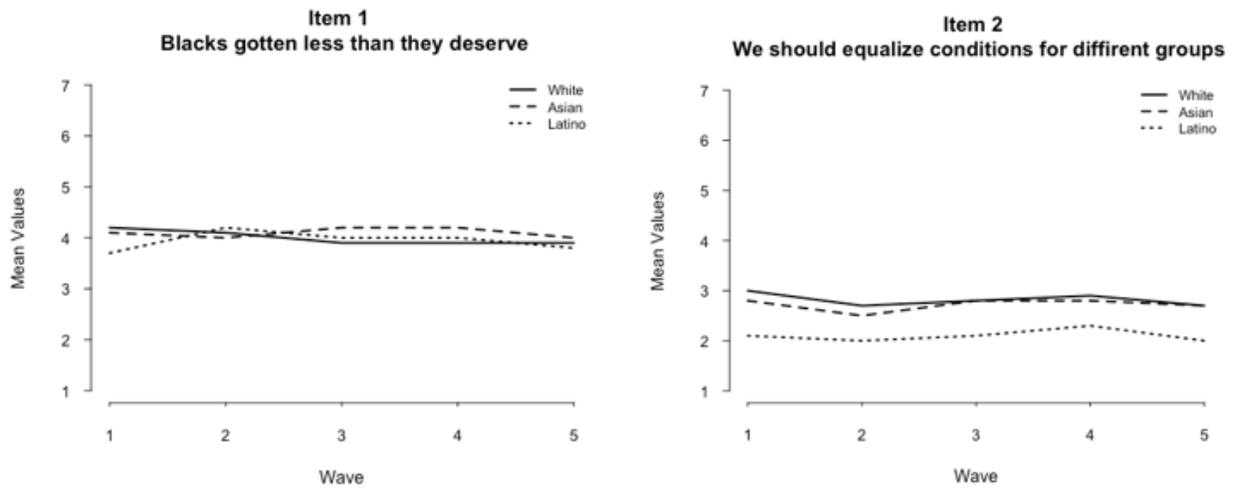

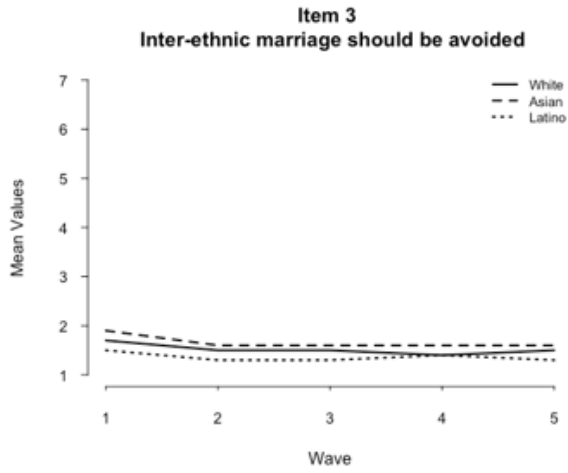
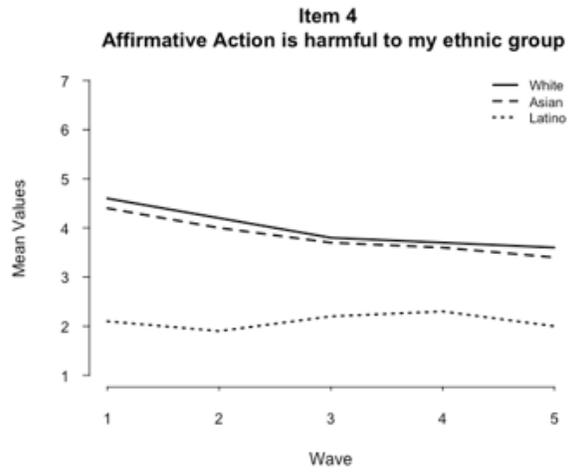
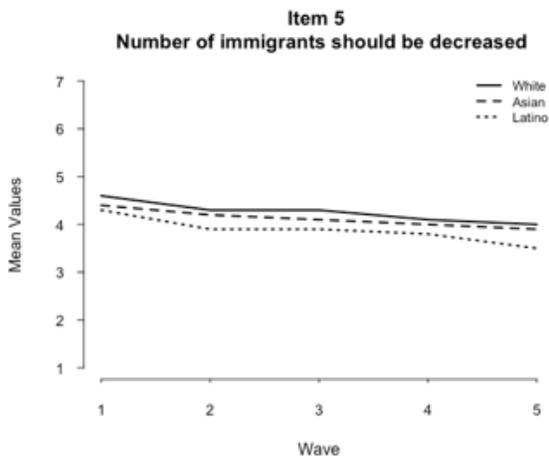
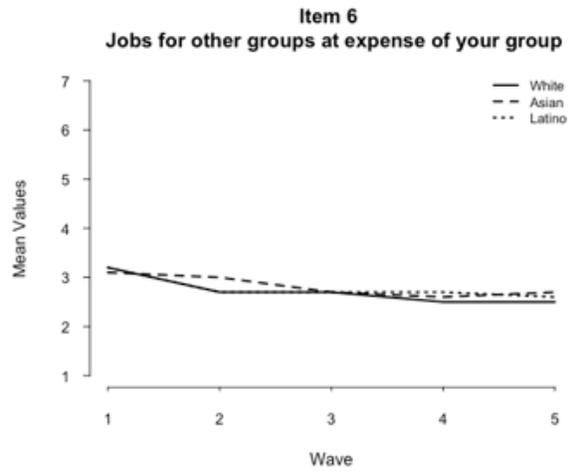

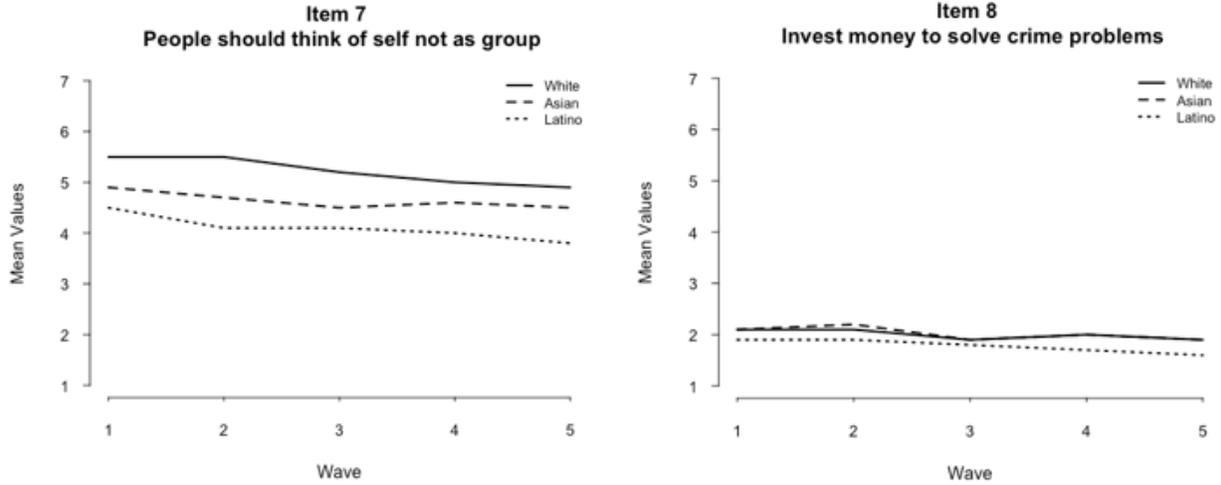

Figure 4 displays the mean attitudes across groups and over five repeated surveys. Respondents rated their agreement or disagreement with eight statements on a scale of 1 (strongly disagree) to 7 (strongly agree): (1) "Blacks have received less than they deserve economically." The attitudes of Whites, Asians, and Latinos were similar and leaned towards the conservative end. (2) "We should equalize conditions for different groups." Asians and Whites held identical attitudes across all waves, while Latinos exhibited more liberal attitudes. (3) "Inter-ethnic marriage should be avoided." The gaps between Asians, Whites, and Latinos were consistent, although all groups expressed liberal attitudes toward inter-ethnic marriage. (4) "Affirmative action is harmful to my ethnic group." Asians and Whites held identical attitudes, which differed from those of Latinos. (5) "The number of immigrants should be decreased." Initially, all groups held conservative attitudes that gradually became more liberal over time. (6) "Jobs for other groups at the expense of your group." Again, all groups tended to have similar attitudes. (7) "People should think of themselves not as a group." Consistent gaps between groups began to emerge. (8) "Invest money to solve crime problems." All groups held similar attitudes towards crime issues. All items were on a 7-point Likert scale and were rescored to assign low values to liberal attitudes.

## 6.1 Determinants of Policy Attitudes

This section aims to explore the factors that influence policy attitudes and investigate whether Whites, Asians, and Latinos share similar predictors. We hypothesize that if these groups' attitudes share a similar set of covariates, they should display comparable magnitudes, directions, and standard errors in the corresponding coefficients. To increase the efficiency of the multivariate analyses, we merged the five-wave panel samples and used an ordered logit for the analysis. The dependent variables were each of the eight items, which were rated on a 7-point scale, ranging from strongly agree to strongly disagree. Some items were reversely scored to ensure that they were all in the same direction, that is, 1—liberal and 7—conservative.

Among the independent variables, partisanship was measured on a 7-point scale, ranging from strong Democrat to strong Republican. Intergroup anxiety was assessed by the item: "I feel uneasy being around people of different ethnicities" (1—strongly disagree, 7—strongly agree). In-group closeness was measured by the item: "Closeness to other members of ethnic group" (1—not at all, 7—very close). First-generation students were measured by the item: "Are you the first person in your family to attend college?" (0—yes, 1—no). Socioeconomic status (SES) was measured by the item: "Family social class position" (1—poor, 8—upper class). Gender was a dichotomous variable (0—female, 1—male). US born was assessed by asking respondents "Were you born in the U.S.?" (0—yes, 1—no). SAT (Verbal) and SAT (Math) were the students' scores for the verbal part (240-800) and math part (200-800) of the SAT.

Figure 5. Determinants of policy preference

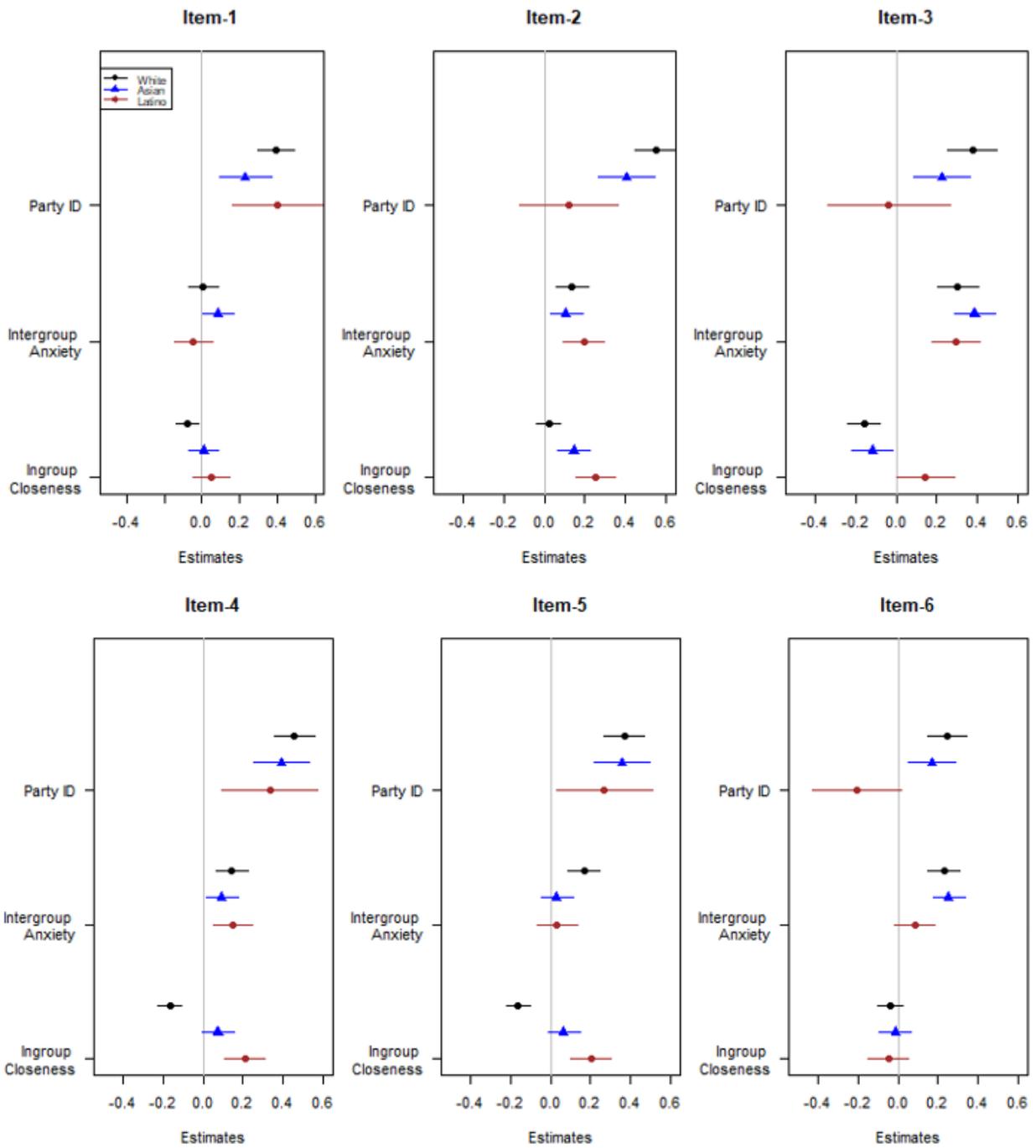

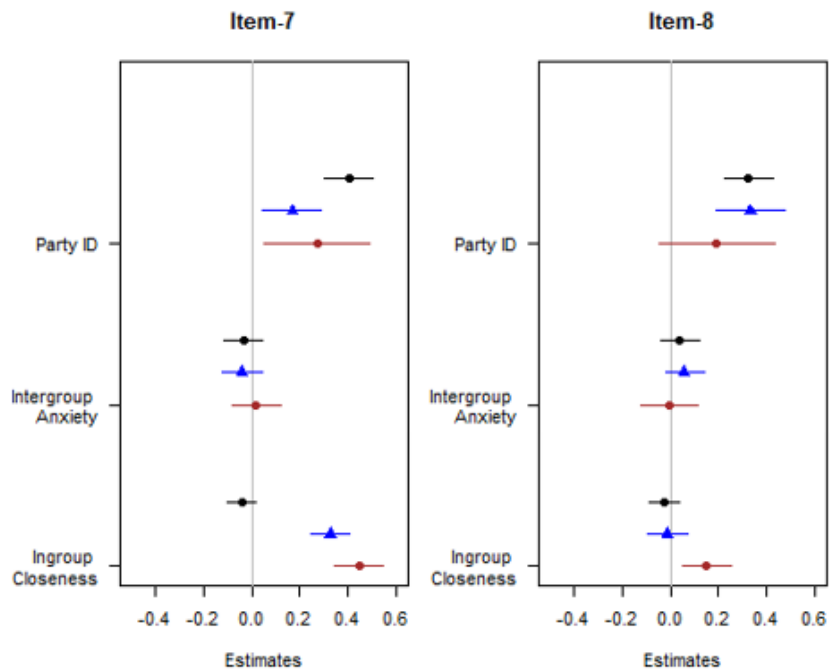

The multivariate analyses suggest that White, Asian, and Latino students had different predictors for item preferences, with little overlap. White students' preferences were most closely aligned with partisanship, with the magnitudes of their partisanship coefficients remaining consistent across all items. While most Asian and Latino students' partisanships were statistically significant, their magnitudes were generally weaker than those of White students, particularly among Latino students. Instead, their item preferences were more strongly related to intergroup relations, group identity cohesiveness, foreign-born status, and the nature of the specific survey questions.

Intergroup anxiety appeared to have a moderate impact on Whites and Asians, but less so than partisanship. For Latinos, the impact varied depending on the specific items. Overall, lower levels of intergroup anxiety were associated with greater support for liberal attitudes on most items. Conversely, the strength of in-group ties was more influential for Latino students, except for items 1 and 7. In general, strong in-group ties tended to push White students towards conservative attitudes in almost all items, whereas for Latino students, strong in-group ties tended to encourage liberal attitudes on most items. Additionally, the regression results (refer to Table A1 in the appendix) showed that certain demographic factors, such as female and foreign-born individuals, were more likely to hold liberal attitudes across all groups.

## *6.2  Assessing Dimensionality of Item Responses*

The regression outputs in Figure 5 suggest that White, Asian, and Latino students had different predictors for their item responses, indicating that their latent attitudes underlying the differential item functioning may differ across groups. To test this hypothesis, we conducted an exploratory factor analysis (EFA) to examine the possible multidimensionality in the item responses. As responses to items are multiply determined, and the items are intended to measure the dominant consideration that is common to all items, understanding the underlying dimensions can provide insights into the differences in the way groups thought about the items. A large eigenvalue in each wave of the survey indicates a strong coherent consideration that accounts for the variation of all items. Figure 6 summarizes the largest eigenvalues in EFA across different groups and measurements. The eigenvalues for White, Asian, and Latino students showed distinct patterns across the five waves of measurements, with White students having the largest eigenvalues in all waves, and Latino and Asian students exhibiting similar and consistently lower patterns compared to White students. These differing patterns of eigenvalues suggest that these groups had different levels of conceptualization of the survey items. For detailed EFA results, see Table A1.

Figure 6. Largest eigenvalues scree plot

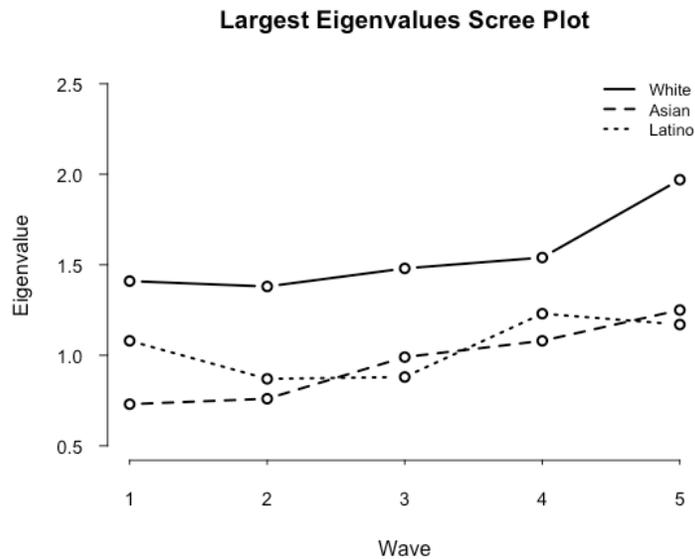

This plot only reports the largest eigenvalues which derived from exploratory factor analysis. See Figure A1 in the appendix for full scree plots.

## 7 Statistical Results of G Theory

This section presents the statistical results obtained through G theory. To better understand the nature of response instability, it is necessary to not only examine the items but also other sources of error variance that can be attributed to specific factors (Feldman, 1989). Therefore, we examined the variance components and their magnitudes for each group and assessed the effects of varying combinations of occasions and items on reliability. It's important to note that in G theory, the variance components are not measured on a universal metric, and their interpretation depends on the relative magnitudes of the different components (Shavelson and Webb, 1991).

Table 1. G theory results for White students

|  | Estimate | Std. Error | 2 2 | 3 3 | 4 4 | 5 5 | 5 6 | 5 7 | 5 8 |
|---|---|---|---|---|---|---|---|---|---|
| Person | 0.450 | 0.059 | 0.450 | 0.450 | 0.450 | 0.450 | 0.450 | 0.450 | 0.450 |
| Occasion | 0.057 | 0.148 | 0.028 | 0.019 | 0.014 | 0.011 | 0.009 | 0.008 | 0.007 |
| Item | 0.328 | 0.275 | 0.164 | 0.109 | 0.082 | 0.066 | 0.066 | 0.066 | 0.066 |
| Person | Occasion | 0.000 | 0.013 | 0.000 | 0.000 | 0.000 | 0.000 | 0.000 | 0.000 | 0.000 |
| Person | Item | 0.275 | 0.026 | 0.138 | 0.092 | 0.069 | 0.055 | 0.055 | 0.055 | 0.055 |
| Occasion | Item | 1.139 | 0.297 | 0.285 | 0.127 | 0.071 | 0.046 | 0.038 | 0.033 | 0.028 |
| Person | Occasion | Item | 1.563 | 0.032 | 0.391 | 0.174 | 0.098 | 0.063 | 0.052 | 0.045 | 0.039 |
|  |  |  |  |  |  |  |  |  |  |
| Reliability | 0.200 |  | 0.460 | 0.629 | 0.730 | 0.793 | 0.808 | 0.819 | 0.827 |
| N |  | 172 |  |  |  |  |  |  |  |

Table 2. G theory results for Asian American students

|  | Estimate | Std. Error | 2 2 | 3 3 | 4 4 | 5 5 | 5 6 | 5 7 | 5 8 |
|---|---|---|---|---|---|---|---|---|---|
| Person | 0.148 | 0.021 | 0.148 | 0.148 | 0.148 | 0.148 | 0.148 | 0.148 | 0.148 |
| Occasion | 0.027 | 0.099 | 0.014 | 0.009 | 0.007 | 0.005 | 0.005 | 0.004 | 0.003 |
| Item | 0.308 | 0.238 | 0.154 | 0.103 | 0.077 | 0.062 | 0.062 | 0.062 | 0.062 |
| Person | Occasion | 0.000 | 0.011 | 0.000 | 0.000 | 0.000 | 0.000 | 0.000 | 0.000 | 0.000 |
| Person | Item | 0.806 | 0.210 | 0.202 | 0.090 | 0.050 | 0.032 | 0.027 | 0.023 | 0.020 |
| Occasion | Item | 0.212 | 0.018 | 0.106 | 0.071 | 0.053 | 0.042 | 0.042 | 0.042 | 0.042 |
| Person | Occasion | Item | 1.547 | 0.026 | 0.387 | 0.172 | 0.097 | 0.062 | 0.052 | 0.044 | 0.039 |
|  |  |  |  |  |  |  |  |  |  |
| Reliability | 0.08 |  | 0.231 | 0.380 | 0.498 | 0.587 | 0.612 | 0.632 | 0.647 |
| N |  | 255 |  |  |  |  |  |  |  |

Table 3. G theory results for Latino students

|  | Estimate | Std. Error | 2 2 | 3 3 | 4 4 | 5 5 | 5 6 | 5 7 | 5 8 |
|---|---|---|---|---|---|---|---|---|---|
| Person | 0.108 | 0.040 | 0.108 | 0.108 | 0.108 | 0.108 | 0.108 | 0.108 | 0.108 |
| Occasion | 0.000 | 0.002 | 0.000 | 0.000 | 0.000 | 0.000 | 0.000 | 0.000 | 0.000 |
| Item | 2.194 | 1.044 | 1.097 | 0.731 | 0.549 | 0.439 | 0.366 | 0.313 | 0.274 |
| Person | Occasion | 0.000 | 0.013 | 0.000 | 0.000 | 0.000 | 0.000 | 0.000 | 0.000 | 0.000 |
| Person | Item | 0.753 | 0.065 | 0.377 | 0.251 | 0.188 | 0.151 | 0.126 | 0.108 | 0.094 |
| Occasion | Item | 0.027 | 0.012 | 0.007 | 0.003 | 0.002 | 0.001 | 0.001 | 0.001 | 0.001 |
| Person | Occasion | Item | 1.292 | 0.041 | 0.323 | 0.144 | 0.081 | 0.052 | 0.043 | 0.037 | 0.032 |
|  |  |  |  |  |  |  |  |  |  |
| Reliability | 0.050 | 0.053 | 0.134 | 0.215 | 0.286 | 0.348 | 0.390 | 0.427 | 0.460 |
| N |  | 85 |  |  |  |  |  |  |  |

Tables 1, 2, and 3 present the G theory statistical results for each group. The total variance is the weighted sum, and the first column on the left side of each table shows the estimated weighted variance components during the first wave of measurement. The first variance component highlighted

in each table is referred to as "person" and represents the true opinion variance of the subjects. The remaining variance components represent sources of error.

Table 1 indicates that the estimated variance attributed to subjects was 0.45, which accounts for 12 percent of the total variance. The estimated variance attributed to items was 0.328, accounting for 8.6 percent of the total variance. In contrast, the estimated variance for occasions was only 0.057, representing 1.5 percent of the total variance. However, the estimated variance for the interaction between persons, occasions, and items was 1.563, accounting for 41 percent of the total variance.

Tables 2 and 3 show the variance components of Asian and Latino students. The estimated variance attributed to subjects was relatively small compared to that of White students, at 0.148 and 0.108. This suggests that Asian and Latino students' opinions on policy issues were relatively consistent compared to those of White students, with most holding liberal attitudes towards item preferences. The estimated variances attributed to items were largest for Latinos, at 2.194, which dominated the measurement error, but smaller for Asians, at 0.308. The estimated variance for occasions was 0.027 for Asians, but almost zero for Latinos, indicating that Latino students' opinions remained highly consistent over time. For both Asians and Latinos, the low variance in subjects and occasions resulted in estimated variances of the interaction between subjects and occasions being close to zero. The estimated variance for Asians and Latinos when subjects interacted with the items were 0.806 and 0.753, respectively. Due to the low variances in subjects and occasions, the estimated variance for the interaction between subjects, occasions, and items was 1.292 for Latinos, which was lower than for Asians (1.547) and Whites (1.563).

We investigated how different combinations of occasions, items, and their interactions affected the reliability across groups using Equation 6 to estimate the reliability by attenuating the error variances of occasions, items, and their interaction. Each combination produced different corresponding variance components. The third column onwards in Tables 1-3 report different combinations of occasions and items, and since there were five waves of survey measurements and eight items, the choices of occasions and items were arbitrary within this limit. We expected to increase opinion reliability by attenuating each error variance component based on the given numbers of items and occasions while keeping the subject's true opinion variance constant. The resulting reliabilities were visualized in Figure 7, which were based on varying combinations of occasions and items reported in Tables 1-3.

Figure 7. Reliabilities on varying waves and items

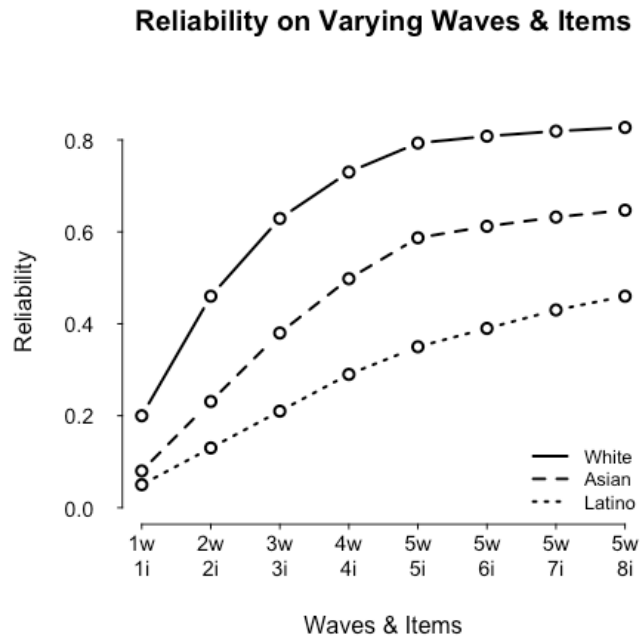

Figure 7 illustrates that White students exhibited the highest reliability, and the magnitude of change was more pronounced as the number of items and measurement occasions increased. Although the true population reliability remains unknown, we utilized bootstrap simulation to estimate an expected reliability of approximately 0.8 (see Appendix Figure A3 for details). With five waves of measurements and five items, the ratio of true opinion variance to total variances was higher for White students, leading to an opinion reliability close to the expected value of 0.8. This reliability level was comparable to that of partisanship and ideology, as well as the reliability of policy attitude scales constructed using common items in Ansolabehere et al.'s (2008) research and the average correlation of knowledge items in Freeder et al.'s (2018) study, where individuals correctly aligned issue positions with candidates or parties. In contrast, due to a smaller ratio of true opinion variance to larger error variance, Asian and Latino students achieved reliabilities of 0.65 and 0.46, respectively, when utilizing five waves of measurements and eight items. These reliabilities slightly exceeded those obtained from the multi-item scale (refer to Table A2 in the Appendix). These findings collectively indicate that in order to achieve opinion reliabilities comparable to those of Whites, Latino and Asian Americans require a greater number of items in their surveys. On the other hand, while multiple waves of surveys can effectively reduce measurement error for Whites due to the larger over-time specific variance, it has less impact on attenuating measurement error for Latinos and Asian Americans, as their over-time specific variance is smaller.

## 8 Discussion & Conclusion

In the context of assessing the preferences of the broader population, public opinion serves as the primary mechanism for political representation, aiming to reflect and represent the policy preferences of ordinary individuals (Campbell et al., 1960; Downs, 1957; Key, 1961). However, survey research in the real world faces a significant challenge in understanding opinion instability. Political scientists have long struggled to identify the factors that contribute to this instability (Achen, 1975; Feldman, 1989; Freeder et al., 2018; Zaller and Feldman, 1992), but the study of over-time and item-specific variabilities across different groups has been hindered by the limited availability of multiple wave panel data and suitable methods.

The analysis using G theory provides valuable insights into the sources of measurement error and the reliability of the measures employed in this study. The examination of measurement error decomposition reveals that a significant portion of the error variance arises from item-specific variance, suggesting inconsistent measurement properties among individual items. The analysis further highlights variations in reliability estimates across different groups, with White students generally displaying higher reliability compared to Latino and Asian American students. This difference can be attributed to discrepancies in item-specific variance, with White students demonstrating lower variance while Latino and Asian American students exhibit larger variance, resulting in greater measurement error. Given that low opinion reliability is largely associated with item-specific variance, it becomes imperative to include a greater number of survey items to mitigate measurement error among minority respondents. Additionally, to enhance overall reliability, careful attention should be directed towards refining and improving the measurement of specific items. Furthermore, the present study reveals that over-time variance, including intervals of repeated measurements, also exhibits group-specific characteristics. Specifically, intervals of measurements tend to generate more error variance for White students, less for Asian American students, and minimal variance for Latino students. This finding underscores the need to consider group-specific factors when designing longitudinal studies and interpreting the reliability of measures over time.

We argue that disparities in political socialization and political conceptualization contribute to variations in levels of item-specificity and attitude stability. The college experience plays a crucial role in shaping students' political socialization, and different groups exhibit distinct characteristics in this process. Disparities in political socialization, acculturation, lived experiences, and issue concerns can lead to divergent perceptions of survey items and considerations among various groups. Factors such as social cognition, cultural competence, social identity, and social stigma may also contribute to these disparities. Consequently, minority students may experience ambivalence when responding to options that align with their partisan identity, social identity, or predispositions. While all groups undergo socialization into partisanship during their college years, the specific factors that shape their predispositions, issue concerns, and social networks influence the level of item specificity required in surveys. Notably, White students typically display a strong alignment between their policy attitudes and partisanship, with the majority of attitude variance stemming from genuine variations in attitudes. In contrast, Asians and Latinos tend to demonstrate more uncertainty and ambivalence toward policy issues.

In conclusion, this study has provided valuable insights into the sources of measurement error and the reliability of measures employed in the research. Despite the limitations associated with student subjects and the use of dated data, the findings carry important implications for survey and public opinion research, particularly concerning minority citizens and cross-ethnic analysis. The study's significance lies in its ability to reveal cross-group, item-specific, and temporal variances that are not easily detectable through regression analysis or the use of multi-item scales alone. By acknowledging these factors and incorporating them into future studies, researchers can improve the measurement quality and reliability of survey instruments.

**Appendix**

Figure A.1. Scree plots of five waves of measurements on survey items

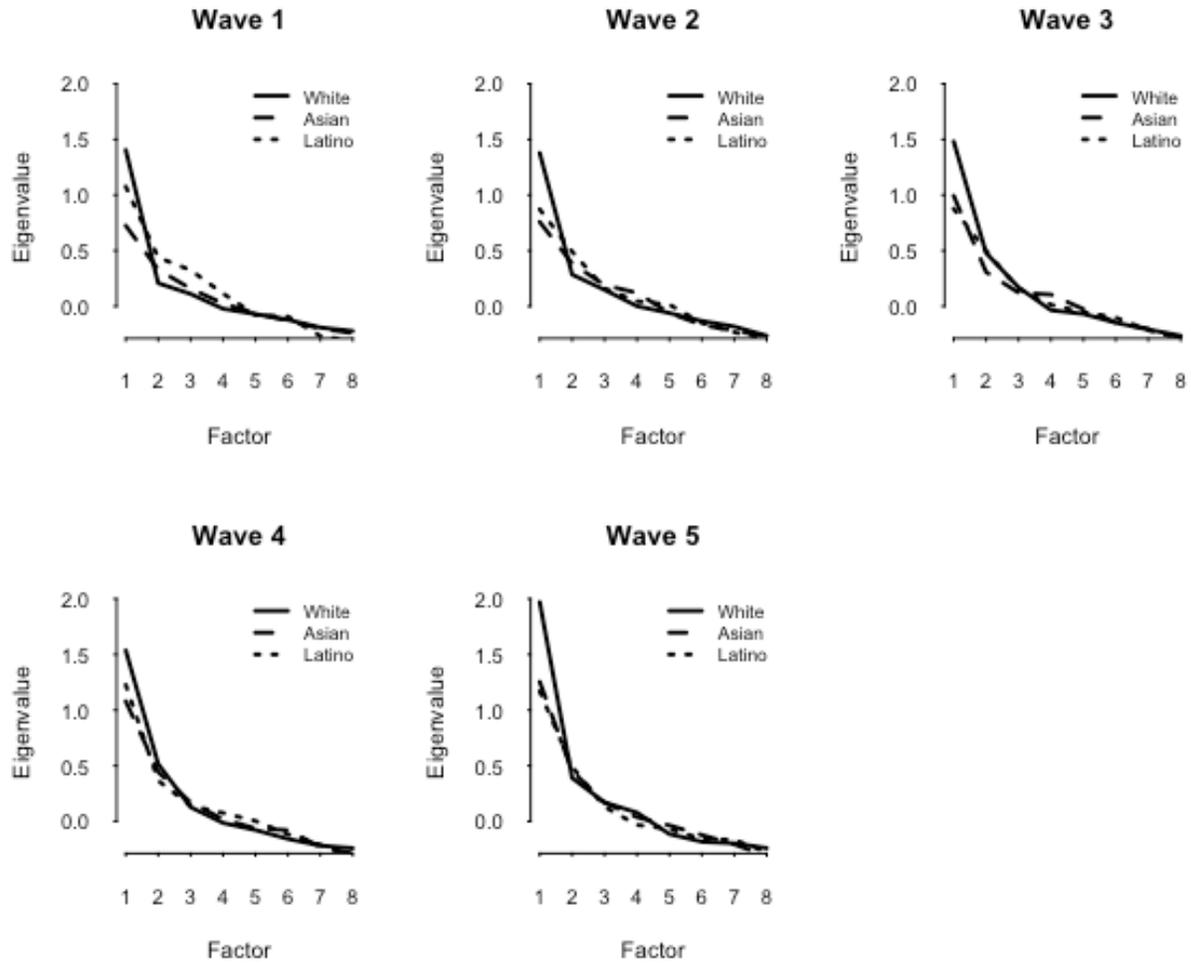

Table A1. Exploratory factor analysis of policy items.

| | White | | | | | | | | | |
|---|---|---|---|---|---|---|---|---|---|---|
| | Wave 1 | | Wave 2 | | Wave 3 | | Wave 4 | | Wave 5 | |
| | Factor 1 | Factor 2 | Factor 1 | Factor 2 | Factor 1 | Factor 2 | Factor 1 | Factor 2 | Factor 1 | Factor 2 |
| Item 1 | -0.399 | 0.254 | -0.280 | 0.299 | -0.384 | 0.246 | -0.377 | 0.338 | -0.426 | 0.270 |
| Item 2 | 0.478 | -0.241 | 0.448 | -0.252 | 0.481 | -0.253 | 0.485 | -0.316 | 0.565 | -0.289 |
| Item 3 | 0.316 | 0.113 | 0.340 | 0.036 | 0.397 | 0.281 | 0.331 | 0.045 | 0.398 | 0.148 |
| Item 4 | 0.535 | 0.142 | 0.580 | 0.178 | 0.579 | 0.182 | 0.598 | 0.212 | 0.634 | 0.221 |
| Item 5 | 0.518 | 0.164 | 0.468 | -0.021 | 0.524 | -0.009 | 0.528 | 0.110 | 0.471 | 0.098 |
| Item 6 | 0.372 | 0.120 | 0.488 | 0.290 | 0.406 | 0.348 | 0.421 | 0.346 | 0.539 | 0.302 |
| Item 7 | 0.354 | 0.025 | 0.209 | -0.022 | 0.341 | -0.209 | 0.368 | 0.114 | 0.433 | -0.039 |
| Item 8 | 0.317 | -0.105 | 0.383 | -0.120 | 0.237 | -0.279 | 0.319 | -0.332 | 0.457 | -0.247 |

| | Latino | | | | | | | | | |
|---|---|---|---|---|---|---|---|---|---|---|
| | Wave 1 | | Wave 2 | | Wave 3 | | Wave 4 | | Wave 5 | |
| | Factor 1 | Factor 2 | Factor 1 | Factor 2 | Factor 1 | Factor 2 | Factor 1 | Factor 2 | Factor 1 | Factor 2 |
| Item 1 | -0.261 | 0.295 | -0.182 | 0.237 | -0.211 | 0.011 | -0.312 | 0.017 | -0.317 | 0.287 |
| Item 2 | 0.572 | -0.207 | 0.522 | -0.001 | 0.447 | 0.116 | 0.522 | 0.222 | 0.514 | -0.139 |
| Item 3 | 0.490 | -0.174 | 0.440 | 0.333 | 0.078 | 0.478 | 0.192 | 0.358 | 0.308 | 0.304 |
| Item 4 | 0.381 | 0.337 | 0.369 | 0.088 | 0.332 | 0.159 | 0.457 | 0.093 | 0.412 | 0.175 |
| Item 5 | 0.298 | 0.013 | 0.291 | -0.294 | 0.382 | 0.041 | 0.526 | -0.128 | 0.450 | 0.228 |
| Item 6 | 0.129 | 0.325 | 0.175 | 0.335 | -0.191 | 0.457 | -0.038 | 0.255 | -0.133 | 0.392 |
| Item 7 | 0.248 | 0.143 | 0.268 | -0.331 | 0.485 | -0.147 | 0.476 | -0.313 | 0.492 | 0.030 |
| Item 8 | 0.359 | 0.198 | 0.223 | -0.090 | 0.316 | 0.011 | 0.327 | 0.015 | 0.281 | -0.243 |

| | Asian | | | | | | | | | |
|---|---|---|---|---|---|---|---|---|---|---|
| | Wave 1 | | Wave 2 | | Wave 3 | | Wave 4 | | Wave 5 | |
| | Factor 1 | Factor 2 | Factor 1 | Factor 2 | Factor 1 | Factor 2 | Factor 1 | Factor 2 | Factor 1 | Factor 2 |
| Item 1 | -0.282 | 0.244 | -0.128 | 0.332 | -0.281 | -0.227 | -0.373 | 0.329 | -0.373 | 0.269 |
| Item 2 | 0.504 | -0.179 | 0.401 | -0.282 | 0.445 | -0.033 | 0.545 | -0.145 | 0.524 | -0.067 |
| Item 3 | 0.399 | 0.201 | 0.340 | 0.171 | 0.481 | -0.260 | 0.365 | 0.354 | 0.496 | 0.326 |
| Item 4 | 0.266 | 0.021 | 0.373 | -0.054 | 0.450 | 0.167 | 0.477 | -0.054 | 0.431 | -0.066 |
| Item 5 | 0.054 | -0.288 | 0.188 | -0.137 | 0.176 | 0.294 | 0.271 | 0.154 | 0.236 | -0.272 |
| Item 6 | 0.262 | 0.280 | 0.453 | 0.179 | 0.397 | -0.121 | 0.383 | 0.313 | 0.488 | 0.277 |
| Item 7 | -0.040 | -0.188 | -0.080 | -0.309 | -0.018 | 0.243 | 0.154 | -0.183 | 0.144 | -0.331 |
| Item 8 | 0.298 | 0.026 | 0.284 | 0.131 | 0.308 | 0.005 | 0.196 | -0.178 | 0.307 | -0.069 |

Note that Table A1 only reported factor 1 and 2, but EFA might also show 3 or 4 factors. However, when factors were more than 2, the factor loadings became weak, which was meaningless to report.

Figure A2. Distribution of survey items across different groups

**Wave 1**

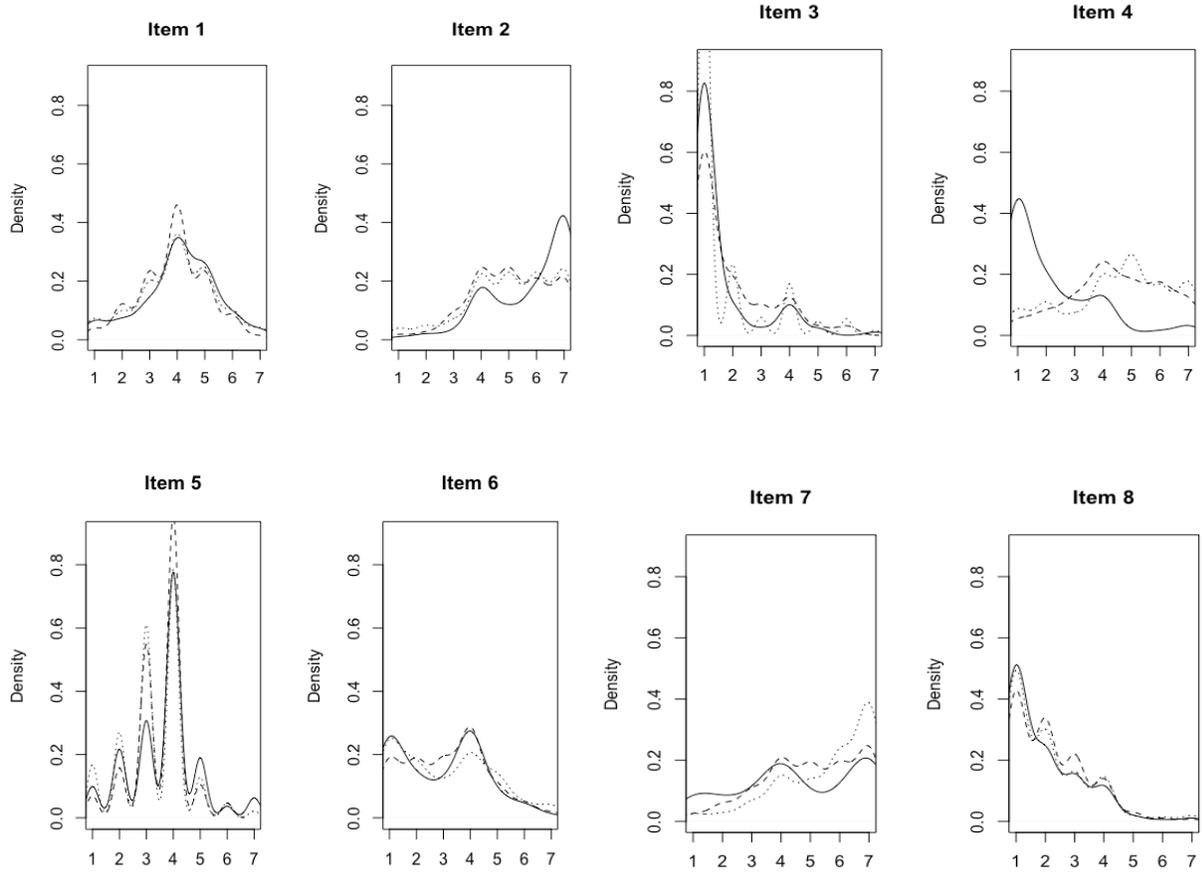

**Wave 2**

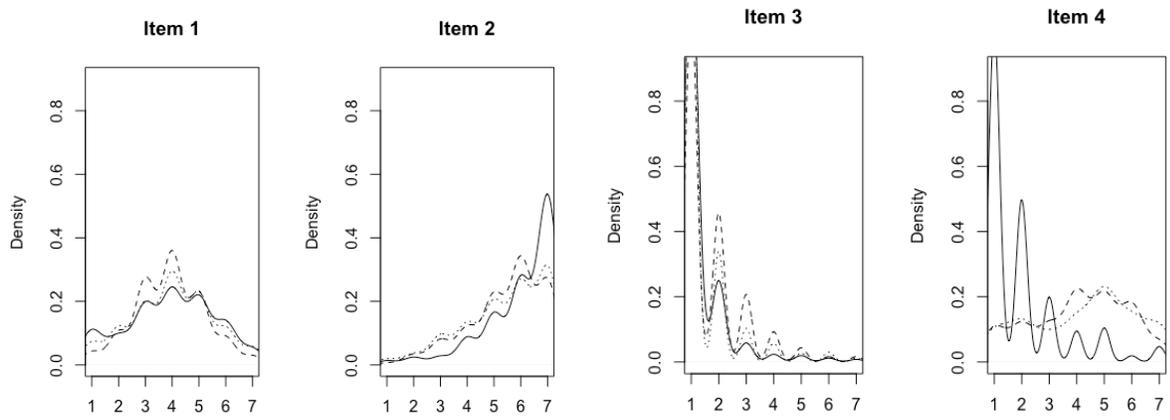

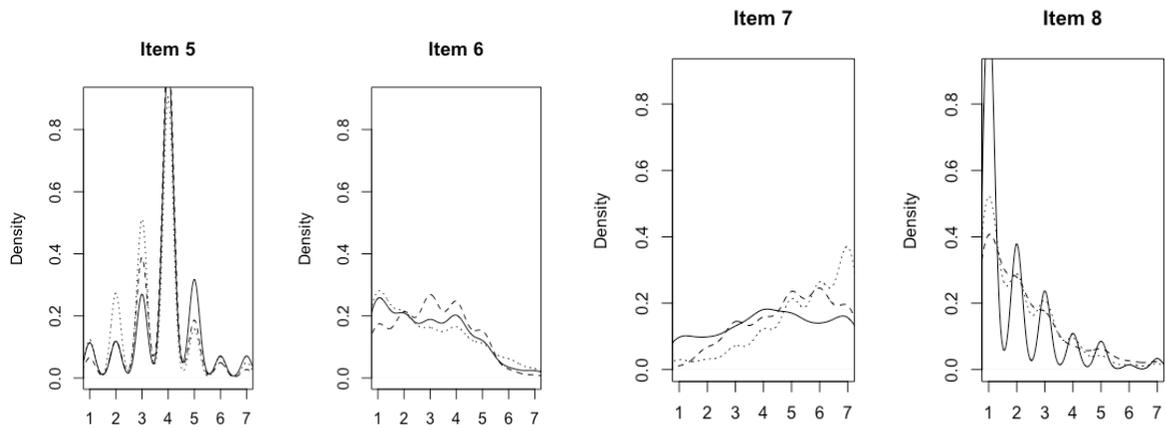

**Wave 3**

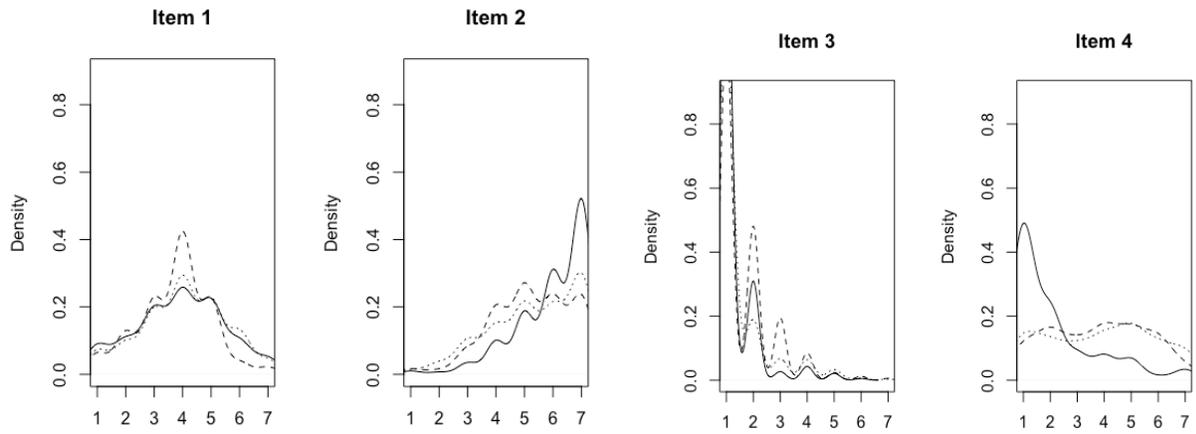

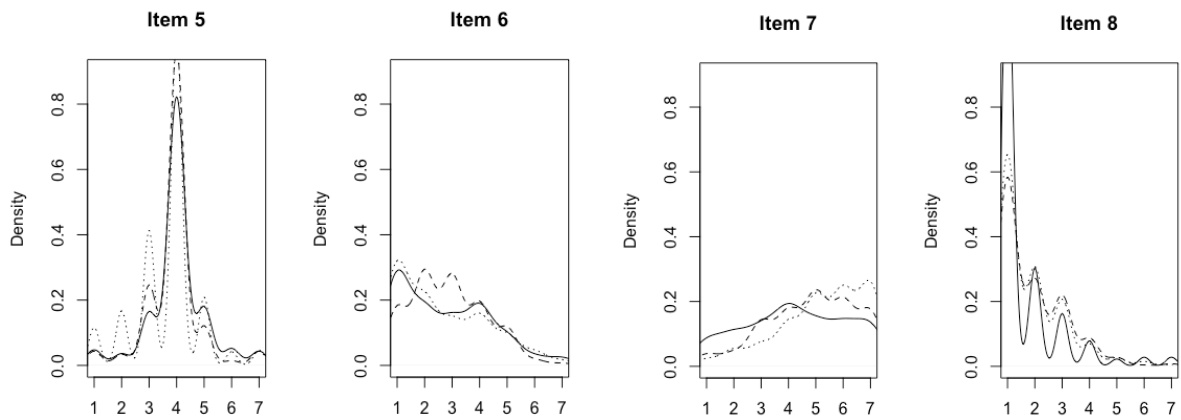

**Wave 4**

| Item 1 | Item 2 | Item 3 | Item 4 |
|---|---|---|---|
| 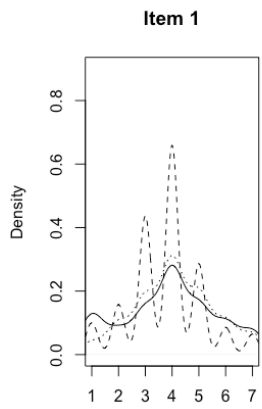 | 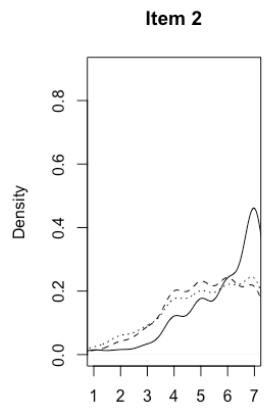 | 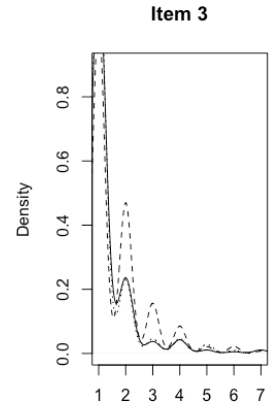 | 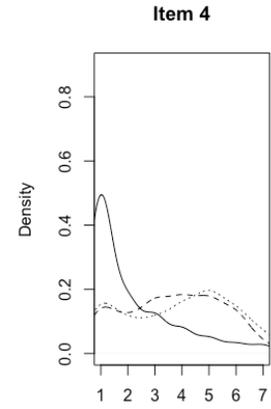 |

| Item 5 | Item 6 | Item 7 | Item 8 |
|---|---|---|---|
| 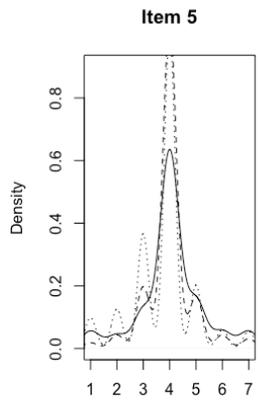 | 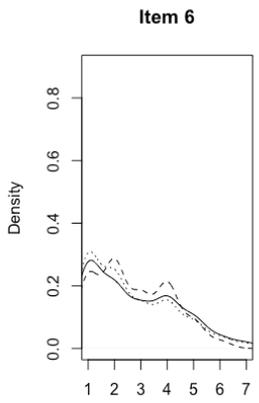 | 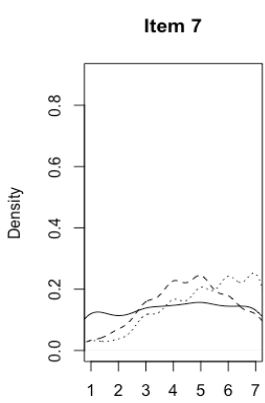 | 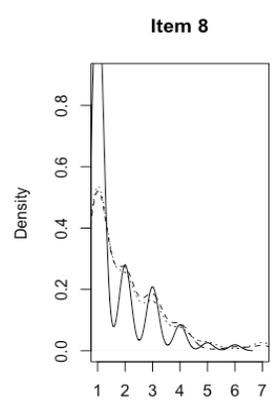 |

**Wave 5**

| Item 1 | Item 2 | Item 3 | Item 4 |
|---|---|---|---|
| 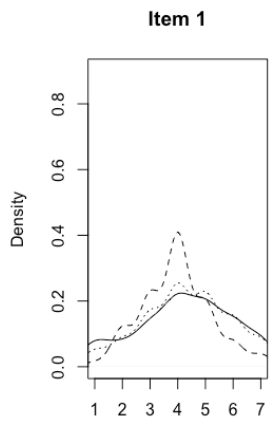 | 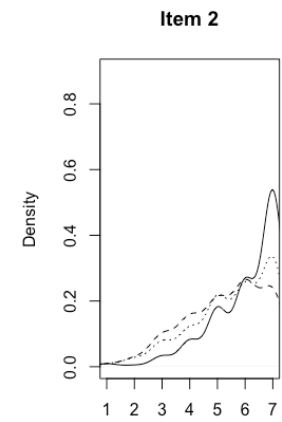 | 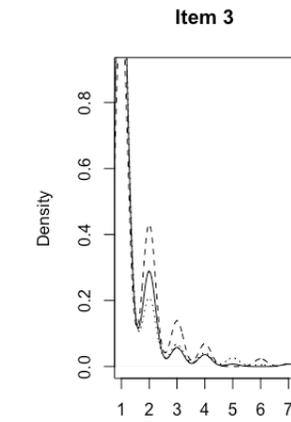 | 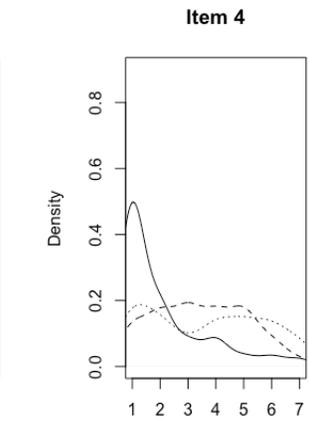 |

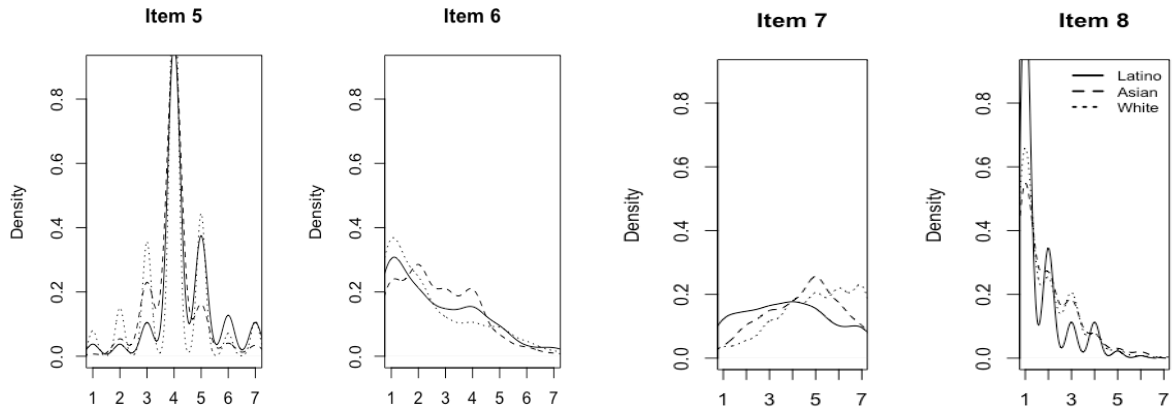

Estimating Expected Reliability

To evaluate the measured attitude reliability, we would need to compare it to the true reliability of the population. Because the true population reliability is unknown, we need to use bootstrap simulation to generate an optimal reliability measure by averaging multiple items. That is, following the logic of equation 5, the more items we average, the higher reliability we can derive. The boxplots in Figure A3 summarizes the simulation results, in which the computer randomly drew $k$ items $k \in \{1, ..., n\}$ from eight survey items with replacement and repeated this process for 1,000 replications for each trial. We calculated the median of $k$ items, as well as their top and bottom 25 percentiles. As the number of items used in constructing scales increased, so did the correlation between them. When $k = 25$, the reliability was about 0.8, which is the theoretically expected reliability that we use as a reference.

Figure A3. Bootstrap simulation of correlations

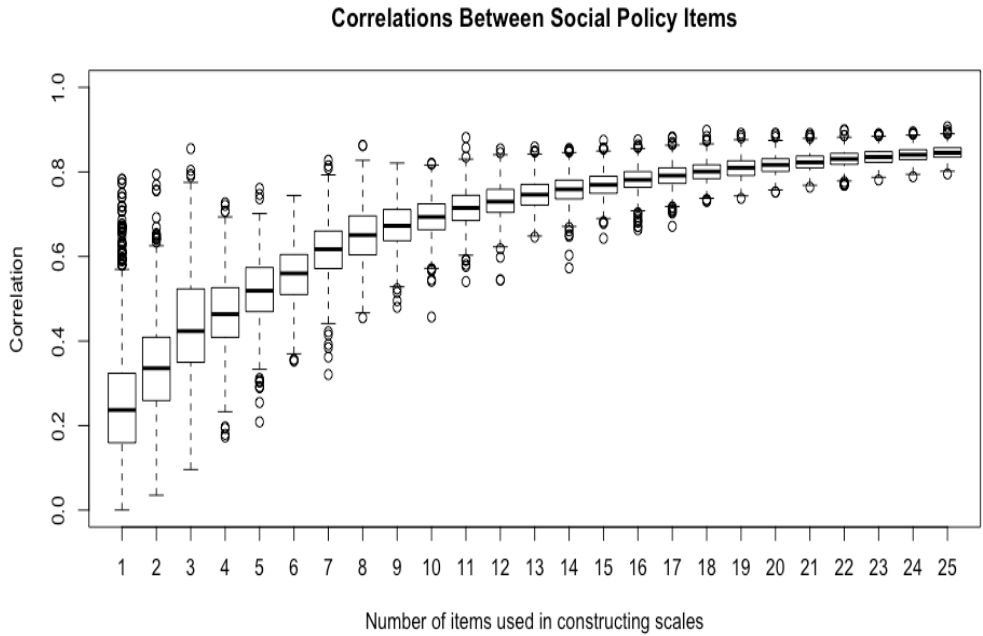

Survey question wording of the items

The following are the eight survey questions that we used to measure attitude stability. The index number is the same as that of the codebook.

Item 1.
Blacks get less attention from the government than they deserve.

Item 2.
We should do what we can to equalize conditions for different groups.

Item 3.
Interethnic marriage should be avoided.

Item 4.
Affirmative action is harmful to members of my ethnic group.

Item 5.
Some people think the number of immigrants who are allowed into the United States should be decreased a lot, some think the number of immigrants should be increased a lot, and some think the number of immigrants should stay the same.

Item 6.
More good jobs for other groups come at the expense of fewer good jobs for members of my group.

Item7.

People should think of themselves first and foremost as an individual American, rather than as a member of a racial, religious, or ethnic group.

The above survey questions are on a 7-points scale, ranging from strongly disagree to strongly agree. Respondents are also allowed to choose 'don't know" or "refused", but these options are coded as missing, and not included in the analysis. The response option for all the questions above was running through the following options:

1. Strong disagree
2. Disagree
3. Somewhat disagree
4. Neither disagree nor agree
5. Somewhat agree
6. Agree
7. Strongly agree

Item 8.
On a scale of 1 to 7, what should we do to solve the crime problems, with 1 meaning investing more money in schools and 7 meaning investing more money in prisons.

Political ideology
On a scale of 1 to 7, how could you describe your general political outlook, with 1 meaning very liberal and 7 meaning very conservative?

Party identification
How would you describe your political party preference? Are you a Democrat, a Republican, an Independent, or do you have some other political party preference? If you are an independent, do you consider yourself closer to Democrat or Republican?

Based on these questions, we construct a 7-point party identification variable
1. Strong Democrat
2. Weak Democrat
3. Leaning more Democrat
4. Neither
5. Leaning more Republican
6. Weak Republican
7. Strong Republican

In-group closeness
How often do you think of yourself as a member of your ethnic group, with 1 meaning not at all and 7 meaning very often.

Intergroup anxiety
I feel uneasy being around people of different ethnicities, on the same of 1 to 7, with 1 meaning strongly disagree and 7 means strongly agree, how much do you agree or disagree with this statement?

Table A1. Determinants of policy preference

|  | Item 1 | | | Item 2 | | | Item 3 | | | Item 4 | | |
|---|---|---|---|---|---|---|---|---|---|---|---|---|
|  | White | Asian | Latino | White | Asian | Latino | White | Asian | Latino | White | Asian | Latino |
| Partisanship | 0.392*** | 0.230*** | 0.401*** | 0.549*** | 0.405*** | 0.121 | 0.374*** | 0.225*** | -0.037 | 0.456*** | 0.392*** | 0.332*** |
|  | (0.05) | (0.07) | (0.12) | (0.05) | (0.07) | (0.12) | (0.06) | (0.07) | (0.15) | (0.05) | (0.07) | (0.12) |
| Intergroup anxiety | 0.006 | 0.085** | -0.047 | 0.136*** | 0.107*** | 0.195*** | 0.303*** | 0.386*** | 0.293*** | 0.143*** | 0.093** | 0.148*** |
|  | (0.04) | (0.04) | (0.05) | (0.04) | (0.04) | (0.05) | (0.05) | (0.05) | (0.06) | (0.04) | (0.04) | (0.05) |
| In-group closeness | -0.082*** | 0.009 | 0.048 | 0.021 | 0.145*** | 0.252*** | -0.158*** | -0.117** | 0.144** | -0.166*** | 0.074* | 0.209*** |
|  | (0.03) | (0.04) | (0.05) | (0.03) | (0.04) | (0.05) | (0.04) | (0.05) | (0.07) | (0.03) | (0.04) | (0.05) |
| Gender | 0.335*** | 0.210* | 0.017 | 0.677*** | 0.390*** | 0.191 | 0.353*** | 0.739*** | 0.746*** | 0.364*** | 0.158 | 0.381** |
|  | (0.09) | (0.11) | (0.15) | (0.09) | (0.11) | (0.16) | (0.12) | (0.13) | (0.20) | (0.09) | (0.11) | (0.16) |
| SAT (Verbal) | -0.004*** | -0.002** | -0.006*** | -0.001* | 0.0005 | 0.001 | -0.003*** | -0.001* | -0.001 | -0.001* | 0.002*** | -0.001 |
|  | (0.00) | (0.00) | (0.00) | (0.00) | (0.00) | (0.00) | (0.00) | (0.00) | (0.00) | (0.00) | (0.00) | (0.00) |
| SAT (Math) | 0.002** | 0.000 | -0.002** | 0.002** | 0.004*** | 0.002** | -0.001 | 0.004*** | 0.000 | 0.003*** | -0.001 | 0.001 |
|  | (0.00) | (0.00) | (0.00) | (0.00) | (0.00) | (0.00) | (0.00) | (0.00) | (0.00) | (0.00) | (0.00) | (0.00) |
| US Born | 0.129 | 0.124 | 0.305 | 0.333* | -0.065 | 0.420** | 0.579*** | 0.058 | 0.766*** | -0.505*** | 0.168 | 0.469** |
|  | (0.18) | (0.11) | (0.20) | (0.18) | (0.11) | (0.20) | (0.22) | (0.12) | (0.24) | (0.18) | (0.11) | (0.20) |
| 1st gen student | -0.101 | -0.017 | -0.336** | -0.124 | -0.289** | -0.199 | 0.473** | 0.013 | 0.21 | -0.498*** | 0.186 | -0.073 |
|  | (0.16) | (0.14) | (0.15) | (0.16) | (0.14) | (0.16) | (0.22) | (0.16) | (0.21) | (0.16) | (0.13) | (0.16) |
| SES | -0.024 | -0.001 | 0.305*** | -0.037 | -0.068 | 0.121 | 0.04 | -0.032 | 0.066 | 0.184*** | 0.135*** | 0.132* |
|  | (0.04) | (0.05) | (0.08) | (0.04) | (0.05) | (0.08) | (0.05) | (0.05) | (0.10) | (0.04) | (0.05) | (0.08) |
| N | 1,727 | 1,099 | 627 | 1,751 | 1,107 | 638 | 1,749 | 1,109 | 637 | 1,738 | 1,105 | 637 |

|  | Item 5 | | | Item 6 | | | Item 7 | | | Item 8 | | |
|---|---|---|---|---|---|---|---|---|---|---|---|---|
|  | White | Asian | Latino | White | Asian | Latino | White | Asian | Latino | White | Asian | Latino |
| Partisanship | 0.369*** | 0.357*** | 0.268** | 0.245*** | 0.169*** | -0.206* | 0.402*** | 0.167*** | 0.271** | 0.323*** | 0.331*** | 0.191 |
|  | (0.05) | (0.07) | (0.12) | (0.05) | (0.06) | (0.11) | (0.05) | (0.06) | (0.11) | (0.05) | (0.07) | (0.12) |
| Intergroup anxiety | 0.168*** | 0.032 | 0.033 | 0.230*** | 0.254*** | 0.084* | -0.035 | -0.039 | 0.017 | 0.04 | 0.059 | -0.004 |
|  | (0.04) | (0.04) | (0.05) | (0.04) | (0.04) | (0.05) | (0.04) | (0.04) | (0.05) | (0.04) | (0.04) | (0.06) |
| In-group closeness | -0.161*** | 0.067 | 0.203*** | -0.04 | -0.013 | -0.047 | -0.042 | 0.325*** | 0.443*** | -0.026 | -0.011 | 0.150*** |
|  | (0.03) | (0.04) | (0.05) | (0.03) | (0.04) | (0.05) | (0.03) | (0.04) | (0.05) | (0.03) | (0.04) | (0.05) |
| Gender | 0.405*** | 0.016 | 0.086 | 0.330*** | 0.307*** | -0.037 | 0.290*** | 0.312*** | -0.173 | 0.470*** | 0.365*** | 0.292* |
|  | (0.09) | (0.13) | (0.16) | (0.09) | (0.11) | (0.15) | (0.09) | (0.11) | (0.15) | (0.09) | (0.12) | (0.17) |
| SAT (Verbal) | -0.005*** | -0.002** | -0.004*** | 0.000 | -0.002*** | -0.002* | -0.002*** | 0.001** | 0.001 | -0.002** | 0.000 | -0.001 |
|  | (0.00) | (0.00) | (0.00) | (0.00) | (0.00) | (0.00) | (0.00) | (0.00) | (0.00) | (0.00) | (0.00) | (0.00) |
| SAT (Math) | 0.002*** | 0.002** | 0.000 | 0.002*** | 0.002*** | 0.001 | 0.002*** | 0.000 | 0.001 | 0.001 | 0.001* | 0.003*** |
|  | (0.00) | (0.00) | (0.00) | (0.00) | (0.00) | (0.00) | (0.00) | (0.00) | (0.00) | (0.00) | (0.00) | (0.00) |
| US Born | 0.157 | -0.536*** | 0.032 | 0.351* | -0.114 | 0.205 | -0.681*** | 0.319*** | 0.674*** | 0.165 | -0.067 | -0.214 |
|  | (0.19) | (0.12) | (0.20) | (0.18) | (0.11) | (0.20) | (0.18) | (0.11) | (0.20) | (0.19) | (0.11) | (0.23) |
| 1st gen student | -0.24 | 0.213 | 0.083 | -0.176 | 0.076 | 0.238 | -0.174 | -0.295** | 0.267* | -0.278 | 0.166 | -0.189 |
|  | (0.17) | (0.15) | (0.16) | (0.16) | (0.14) | (0.15) | (0.17) | (0.13) | (0.15) | (0.17) | (0.14) | (0.17) |
| SES | 0.242*** | -0.077 | 0.193** | 0.028 | 0.044 | 0.116 | 0.016 | -0.079* | 0.02 | -0.011 | 0.041 | 0.152* |
|  | (0.04) | (0.05) | (0.08) | (0.04) | (0.05) | (0.08) | (0.04) | (0.05) | (0.07) | (0.04) | (0.05) | (0.09) |
| N | 1,728 | 1,092 | 636 | 1,742 | 1,106 | 635 | 1,739 | 1,109 | 638 | 1,742 | 1,102 | 638 |

Note: *p<0.1; **p<0.05; ***p<0.01

## Policy Attitude Reliability Using Multi-Item Scale

Table A2 presents the reliability of attitudes towards race for White, Asian American, and Latino individuals, based on eight policy attitude items measured over five waves, using the multi-item scale presented in equation 5. Table A2 provides a summary of the reliability measures, where $\rho_{wi}$ represents the reliability of the eight items in the $i$th wave of measure ($i = 1, \cdots, 5$), and $\bar{\rho}$ is the average of the five repeated measures.

Table A2. Reliability using multi-item scale

|        | $\rho_{w1}$ | $\rho_{w2}$ | $\rho_{w3}$ | $\rho_{w4}$ | $\rho_{w5}$ | $\bar{\rho}$ |
|--------|-------------|-------------|-------------|-------------|-------------|--------------|
| White  | .59 (.03)   | .63 (.02)   | .65 (.03)   | .67 (.02)   | .74 (.02)   | 0.66         |
| Asian  | .36 (.06)   | .48 (.04)   | .60 (.04)   | .64 (.03)   | .67 (.03)   | 0.55         |
| Latino | .46 (.04)   | .45 (.04)   | .43 (.04)   | .35 (.05)   | .45 (.04)   | 0.43         |

Note: standard errors are in parenthesis

Table A2 presents that, on average, White students had the highest reliability across all repeated measurements, with an average of 0.66. In contrast, Asian American and Latino students had consistently lower average reliabilities of 0.55 and 0.43, respectively, compared to their White counterparts.